# Constraining the global mean surface temperature during 1850-1880 with new statistical physical model

Qingxiang Li[1,2#*†], Zichen Li[1†], Xuqian Li[1], Zengyun Hu[2], Aiguo Dai[3], Wenjie Dong[1#], Boyin Huang[4], Zhihong Jiang[5], Panmao Zhai[6], Tianjun Zhou[7], Phil Jones[8]

[1] *School of Atmospheric Sciences and Guangdong Province Key Laboratory for Climate Change and Natural Disasters, SUN Yat-Sen University; Zhuhai, 519082, China*

[2] *Research Center for Ecology and Environment of Central Asia, Chinese Academy of Sciences; Wulumuqi, 830011, China.*

[3] *Department of Atmospheric & Environmental Sciences, University at Albany, SUNY, Albany; New York, 12222, USA.*

[4] *National Centers for Environmental Information, NOAA; Asheville, E/NE42, USA*

[5] *School of Atmospheric Sciences, Nanjing University of Information Science & Technology; Nanjing, 210044, China.*

[6] *Chinese Academy of Meteorological Sciences, CMA; Beijing, 100081, China.*

[7] *Key Laboratory of Atmospheric Sciences and Geophysical Fluid Dynamics, Institute of Atmospheric Physics, Chinese Academy of Sciences; Beijing, 100028, China.*

[8] *Climatic Research Unit, School of Environmental Sciences, University of East Anglia; Norwich, NR4 7TJ, UK.*

# *Southern Laboratory of Ocean Science and Engineering (Guangdong Zhuhai), Zhuhai, China*

*Correspondence*: liqingx5@mail.sysu.edu.cn

† These authors contribute to the paper equally.

**Abstract**

As IPCC ARs stated, global warming is estimated based on the average from 1850 to 1900 (global average temperature of pre- industrialization estimated from relatively sparse observations). Given the impossibility of massive increasing observation data in the early stages, accurately constraining this baseline has become an unresolved issue. Here we developed a new statistical physical model to quantify the contribution of external forcings to global warming as a "deterministic trend" of the surface temperature series (instead of as non-stationary processes that yield a stochastic trend) and constrained the reconstruction of the early time series (1850-1880). We find that the existing datasets slightly overestimated the temperature anomalies in this period, thus the speed of global warming since pre-industrialization is still underestimated.

*Introduction.*—The Intergovernmental Panel on Climate Change's (IPCC) Sixth Assessment Report (AR6) gave a uniform warming estimate (1.09 °C) between the last decade (2011-2020) relative to the period of global pre-industrialization [1] (the average temperature from 1850 to 1900 is often used as a representative), it was close to the 1.5 °C temperature target proposed by the 2015 Paris Agreement. In contrast, the warming was only 0.78°C in AR5 [2], which included the actual temperature increase in recent years (i.e., the difference between the global average temperatures in 2011-2020 and 2003-2012 was about 0.2°C), as well as the changes caused by the reconstruction of the new dataset and the improvement of the calculation method of the warming trend (about 0.1°C). Therefore, it is urgent to prevent and slow down further global warming.

AR6 also stated that all observational datasets gave somewhat different ranges of warming (95% confidence interval of 0.95 to 1.20 °C) [3]. From a comparison of the Global Mean Surface Temperature (GMST) time series from 1850 to 2022 based on different datasets [4-8] (FIG. 1, see also SM for the details), this difference is mainly reflected by the pre-1900 data [9-11]. Brohan et al. and Li et al. pointed out that the most prominent uncertainty component in the global and regional Surface Air Temperature (SAT) anomaly series is the under-sampling of observations [12,13]. Moreover, quality control and homogenization of early data are quite difficult due to the lack of corresponding reference data. Unfortunately, it is currently almost impossible to massively increase observations at an earlier stage (generally before 1950 and especially before 1900). Since global warming is estimated from 1850-1900 averages, as in the latest IPCC reports, constraining the 19th-century temperature baseline for global warming has become an outstanding issue [14,15].

However, we should also see that, despite the above-mentioned problems/ shortcomings, the accuracy of global Surface Temperature (ST, which is merged with Land Surface Air Temperature (LSAT) and Sea Surface Temperature (SST)) observations has reached such a high level that there is no disagreement to say that it is the "most accurate" of the Essential Climate Variables (ECVs) estimates [16]. On the other hand, with the increased research on the attribution of anthropogenic factors in recent decades [17,18], scientists have largely clarified the physical mechanism of external forcing affecting climate warming. Such studies are based on two main aspects: 1) observational data, including climate factor variability and estimate of external forcing; 2) understanding of the contribution of external forcing based on a physical basis and the estimation of the Internal Variability (IV, the portion of the observed series subtracted from the external forcing signal). On this basis, if we can establish a mathematical/statistical model with a more explicit physical understanding that directly decomposes the contribution of human activities (external forcing) to global and regional warming in the ST series, is it possible to obtain the more accurate reconstruction of the early SAT/ST series?

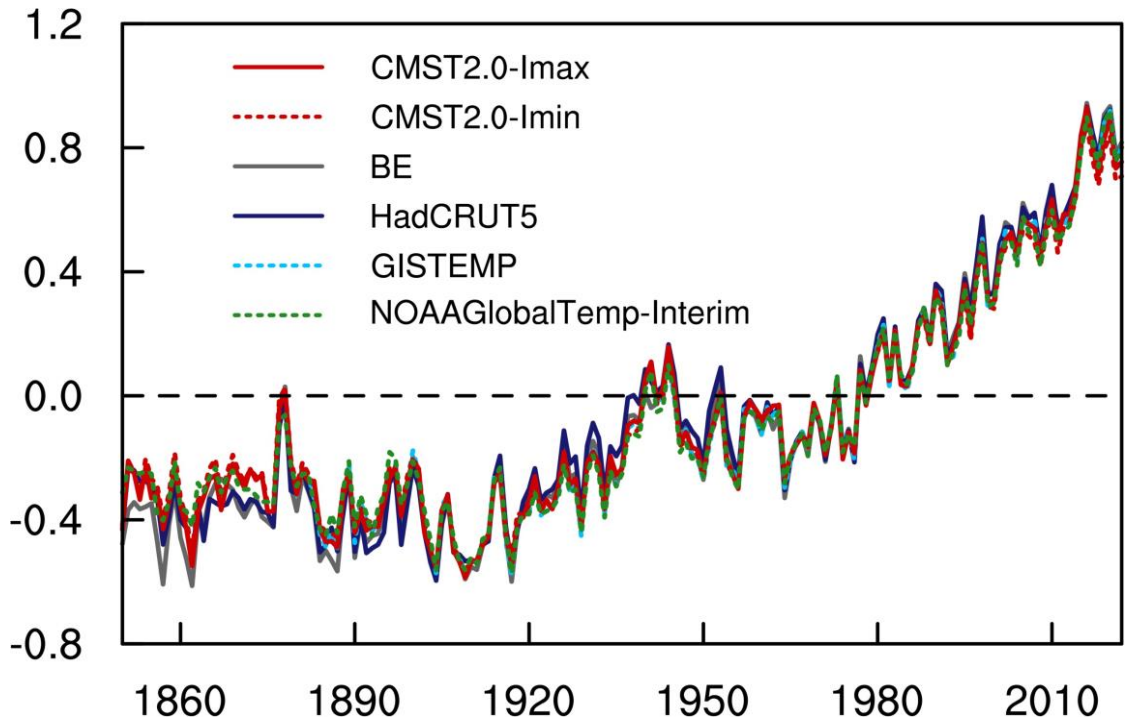

FIG. 1. Comparison of GMST time series derived from several temperature datasets adopted in IPCC AR6 and CMST2.0 (Imax and Imin) during 1850-2022.

*Modeling of the ST series: from FBM to multi-linear response process.—* Mandelbrot and Van Ness proposed the Fractal Brownian Motion (FBM) model and establish a complete and self-consistent system of study [19]. For time series with a long memory, one can study the influence of their long memory on the variation of the time series. FBM is perfectly applicable to many problems in natural sciences, engineering, and socio-economic statistics [20,21]. If the atmospheric motion is considered in separate time steps, it is consistent with the classical random walk process [22]. According to the random walk (or FBM) theory, if the ST series is recognized as a random walk model with "drift" (characterized by a differential stationary process, FIG. S1), the warming trend of the GMST can also be regarded as the sum of "drift" "a" and the stochastic trend "$\sum_{t=1}^{n} \mu_t$" (assuming the initial value $y_0 = 0$, $\mu_t$ is a

white noise) in Eq. (1):

$$y_t = a + y_{t-1} + \mu_t \qquad (1)$$

From a statistical point of view, such time series as "random walk" (FBM) cannot be extrapolated (predicted) precisely because it contains "stochastic" trends [23].

Later, Malamud and Turcotte stated that any time series can be perceived to be composed of three parts: a trend term, one or more periodic terms, and a stochastic term [24]. Based on their classification, many series would have periodic components, and the stochastic component is the fluctuations not included in either the trend or periodic components. The trend component here is a long-term increase or decrease in the series and is regarded as a "deterministic trend". Thus, the time series can be expressed as a deterministic trend "$\beta t$" plus a stationary stochastic process with mean zero (trend-stationary process) (Eq. (2)):

$$y_t = \alpha + \beta t + \epsilon_t \qquad (2)$$

$$\emptyset(L)\epsilon_t = \theta(L)u_t; u_t \sim i.i.d.(0, \sigma_u^2)$$

where $\alpha$, $\beta$ are fixed parameters, $L$ is the lag operator, and $\emptyset(L)$, $\theta(L)$ are polynomials in $L$ that satisfy the conditions for stationary and invertibility. Strictly speaking, this type of time series cannot either be accurately predicted because it contains a random term, although its contribution to the "deterministic trend" is much smaller. The exception occurs when the regression equation satisfies the Best Linear Unbiased Evaluator (BLUE, i.e., the residual of the regression equation is considered a white noise) when there are no random terms.

Nelson and Plosser pointed out that the time series with red noise properties can be better modeled by the random walk process with "drift" (this characteristic is similar to the climate time series), when they study the macroeconomic time series for the United States [20]. According to their comparison, the "trend" of this random walk (sum of stochastic term and deterministic trend term, i.e., $\alpha + \sum_{t=1}^{n} \mu_t$) is, under certain conditions (the fluctuations around the "deterministic trend" are so highly autocorrelated as to be indistinguishable from non-stationary series themselves in realizations as long as one hundred years), essentially equivalent to a long-term trend (deterministic trend) dominated by external forcing. If this "deterministic trend" can be quantified successfully, the model is quite advantageous for global or even regional climate change simulations and predictions [25]. The above conditions are very easy to meet for a temperature change series.

Thus, how to find and express the "deterministic trend" of the climate change series becomes the key to the problem. For condensed molecular systems, it has long been found that the linear response matrix (e.g., conductivity) can be expressed in terms of the Green-Kubo formulation as an improper time integral of the quadratic correlation function in the system [26,27]. Ruelle extended the Green–Kubo formulation to describe a "sufficiently chaotic" nonlinear dynamic system for parameter statistical steady-state response to changes in parameters [28,29]. Lorenz stated that "Climate is deterministic and autonomous, but highly nonlinear; The trajectories diverge exponentially, forward asymptotic PDF is multimodal" [30]. Thus, he believes that the climate cannot be predicted. However, Hasselmann proved that "Climate is stochastic and noise-driven, but quite linear; The trajectories decay back to the mean, forward asymptotic PDF is unimodal" from a stochastic climate model [31]. Based on this, he believed that the climate can be predicted by at least 50%. Later, this linear response feature was more widely applied to global warming studies to calculate the response of the mean temperature to external forcings such as greenhouse gases in the atmosphere: climate sensitivity is defined as the linear response of the long-term mean value to sufficiently small changes in external forcing [32], which is expressed mathematically as an appropriate function of the system state or derivative of other functions to the bifurcation parameter [33].

Our previous studies proposed a more reasonable quantitative decomposition combination model using the radiation forcing factors [34] as the independent variables (see SM for the details), and successfully decomposed the mean temperature series into external forcing responses and multiple periodic terms ($\epsilon_t$), which includes the natural variability and a stochastic term with persistence represented by Autoregressive Integrated Moving Average (ARIMA) model items in Eq. (3) [9,35,36]. Where, $x_1(t)$, $x_2(t)$ are the total natural and anthropogenic forcing (both are the functions of time t), respectively, $\epsilon_t$ is the residual series of the Multivariable Linear Regression (MLR) model, $a_0, a_1 and a_2$ are the regression coefficients evaluated by the Ordinary Least Square (OLS) method. In this case, Eq. (3) will need to be combined with ARIMA models [16], thus the term "$\beta t$" in Eq. (2) is replaced by the term "$a_1 x_1 + a_2 x_2$" in Eq. (3). The previous comparison indicates that this decomposition avoids the uncertainty of the CMIP6 model in the historical simulation of regional temperature series [35].

$$y_t = a_0 + a_1 x_1(t) + a_2 x_2(t) + \epsilon_t \qquad (3)$$

Although the two kinds of "trends" are estimated here in general agreement, in theory, the climate change trend "$a_1 x_1(t) + a_2 x_2(t)$" in Eq. (3) is still not a fully "deterministic trend" because the trend is also affected by the persistence of the climate series [35]. Referring to the idea of the classic fingerprint attribution of climate change [18,37], and using generalized linear regression, "$a_1 x_1(t) + a_2 x_2(t)$" in Eq. (3) is substituted for the global ST response to natural and anthropogenic forcings [35], and the modeling has a traceable physical basis. However, if more combinations of external forces are considered, Eq. (3) will appear ill-conditioned due to the obvious multicollinearity between different external forcings themselves (Table S1). At this time, more reasonable regression modeling needs to be considered. According to Li [9] and Qian et al. [36], a Partial Least Squares Regression (PLSR, see SM for the detail) [38] method has good results in dealing with this multicollinearity. Based on the above, a new regression (Eq. (4)) has been set up. Here $y_t$ is a dependent variable, and $x_i(t)(t = 1,2,...,n)$ are independent variables, $\varepsilon_t$ is the residual. If the residual is stationary, then the whole regression equation can be regarded as a trend stationary process, then the trend term "$\sum_{i=1}^{n} a_i x_i(t)$" is the deterministic trend, and $y_t$ has the potential for accurate prediction/ extrapolation. In addition, we also noticed that there is no significant correlation between all external forcing factors (mainly trend changes, FIG. S2) and important IV factors (such as IPO, AMO, etc., mainly cyclical changes) (Table S1), which seems to ensure that the decomposition of external forcing response and IV are not easily phase mixed.

$$y_t = a_0 + \sum_{i=1}^{n} a_i x_i(t) + \varepsilon_t \qquad (4)$$

Therefore, if we assume that the multivariate linear equation consisting of various radiative forcing factors is stable during instrumental observations, then we can more accurately decompose the climate time series into external forced "signal" and the IV of the climate system, which is manifested as the periodicity in the high- and low-frequency domains. If the climate series has significant autocorrelation, non-stationarity, etc., then the decomposition model needs to be supplemented with a stochastic process. Based on the above, the external forcing "signal" of a time series (e.g., GMST) can be more accurately expressed by the linear response of various external forcings. We here name this model as the Linear External Forcing Response Model (LEFRM). For early stages, we can extend the external forcing components from the observational data that seems to have small biases and relatively reliable variations, and therefore reconstruct the observations in the early period, since the effect of IV on the trend seems to be negligible for the long-term variations.

*Benchmarking the LEFRM with climate model outputs.* —We first tested the LEFRM with the commonly used earth system models' simulations. For GMST, we can always estimate the response as a linear function of the external forcings on the top of the atmosphere [36]. Climate model outputs have always been considered useful tools for understanding underlying causal mechanisms. Here what we need to know is whether the net response can be expressed as a linear combination of the individual forcings if we put all the forcings together in a climate model. We compared the results of the external forcing "signal" by the ensembles of two sets of models output over two time periods (1850-2017 and 1880-2017) (FIG. 2). Two ensemble methods were used here: 1) ensemble of the large sample set of NCAR CESM2 Large Ensemble (LENS2) [39] and 2) multi-model ensemble(MME) of the output set of CMIP6 all-forcing large sample models (Table S3) [25] (see SM for the details). Regardless of the situation, our model explains about 98% of the variance contribution of external forcing changes (FIG. 2a-d). For different model ensemble sets and in different periods, there is no statistically significant difference in the standardized coefficient (used to represent the relative contributions of the different forcings) of each external forcings in the model (FIG. 2e). It suggests that this model can effectively decompose the contributions of each external forcing in the ST change series as climate system models do.

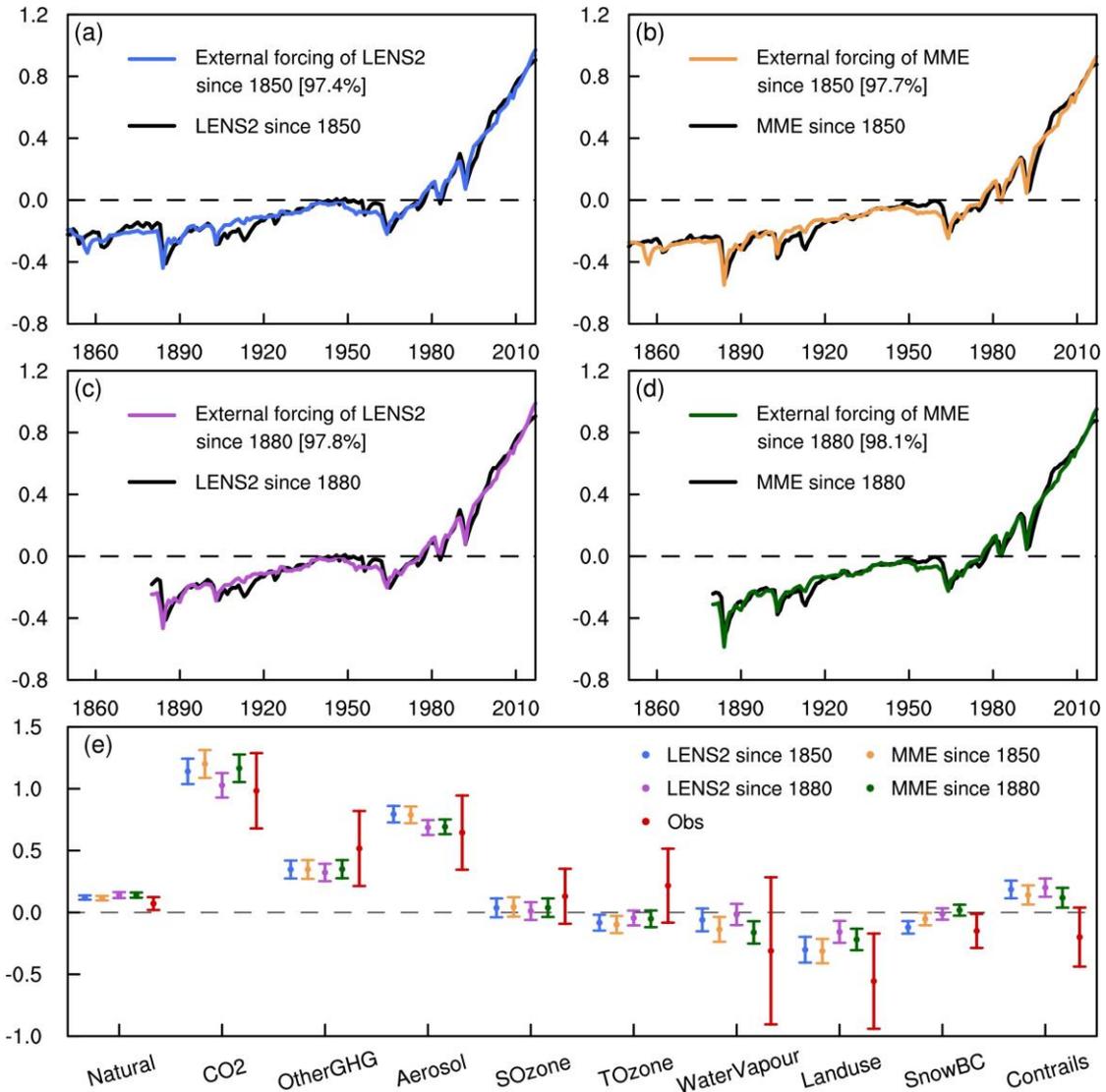

FIG. 2. Decomposition results through the LEFRM. The external forcing of CESM2 Large Ensemble (LENS2) in 1850-2017 (a), multi-model ensemble (MME) of CMIP6 all-forcing simulations in 1850-2017 (b), LENS2 in 1880-2017 (c), and MME in 1880-2017 (d) (The numbers in square brackets represent the explained variance of the equation); (e) the standardized coefficients of the PLSR equations for the above cases (±1σ) and Observation (±1.96σ).

*Response of each external forcing.* —According to Li et al. [9], the differences between the series since 1880 are relatively small (also from the comparison of the trends in Table 1). Due to the reliability of the LEFRM for external forcings decomposition, we trained (fitted) the model with the average series (AVE) of five datasets (the latest upgraded CMST2.0, including Imax and Imin variant and four other datasets (HadCRUT5, BE, NOAAGlobalTemp-Interim, GISTEM4) used by AR6 [1]) from 1880 to 2017. Using the LEFRM (which explains 90.5% of the total variance of

AVE of all datasets) and the radiative forcing factors (FIG. S2) for 1850-2017, the external forcing component was extended forward to 1850. We then calculated the IV component for 1850-1880 and combined it with the external forcing component for the same period. Finally, the reconstructions of the observed early series were obtained.

Taking CMST2.0-Imax (see SM for the details) as an example, FIG. 3 shows the external forcing and IV of the reconstructed series, as well as the variations of GMST responses to all the forcings (with the range of ±1σ). The model results show that $CO_2$, Land Use, and other GHG contribute the most parts of the warming during the whole period of 1850-2017, and Aerosols contribute to the main cooling during this period (FIG.2e and FIG. 3c). The contributions from the rest of the external forcing are relatively small or insignificant. The above understandings are broadly consistent with the attribution results from the classical optimal fingerprint (OFP) method [33,40]. For the radiative forcing factors that contribute more to the temperature change, the standardization coefficients of the equations do not differ significantly between the observations and models in the four cases (FIG. 2e), demonstrating that the responses of temperature to the above factors are relatively stable. The residual of the LEFRM can be regarded as the IV of the global ST change, which indicates little contribution to the long-term warming trend but will have influence at a shorter time scale (FIG. 3a-b).

Further, we utilized the DAMIP models output (Table S3) to analyze the contribution of the individual forcing (including CO2, Aerosol, Natural forcing, etc.) and to compare them with the results the LEFRM achieves (FIG. 3c). Besides, we compared the decomposition results of our model to those of the 3 variables (GHG, AER, and NAT) attribution based on the OFP method [38] (Table S5). Although there are slight differences in the contribution of the individual forcings (Table S4 and FIG. S3, this may be also related to the fact that there are still some multicollinearities between some external forcings (CO2, Aerosol, BC_snow, etc.) and the IV (mainly AMO) after the removal of their linear trends in our model (Table S2)), the decomposition results of our model largely agreed with the separation of the individual forcings contribution and the final attribution of OFP. It is worth noting that in Gillett et al., well-mixed human-emission external forcing, such as ozone, is incorporated into the GHG forcing [41]. Thus, its GHG and AER have included all anthropogenic forcing so that the resulting warming is almost identical to the observed warming. In contrast, neither the GHG nor the AER forcing in our model included OZone, WaterVapour, Landuse, and SnowBC, and the resulting forced response of GMST was slightly lower than the response to all forcings. Considering this, the LEFRM is accurate for the decomposition of the GMST response to different external forcing, thus exhibiting good performance in reconstructing the external signals in early-stage observation.

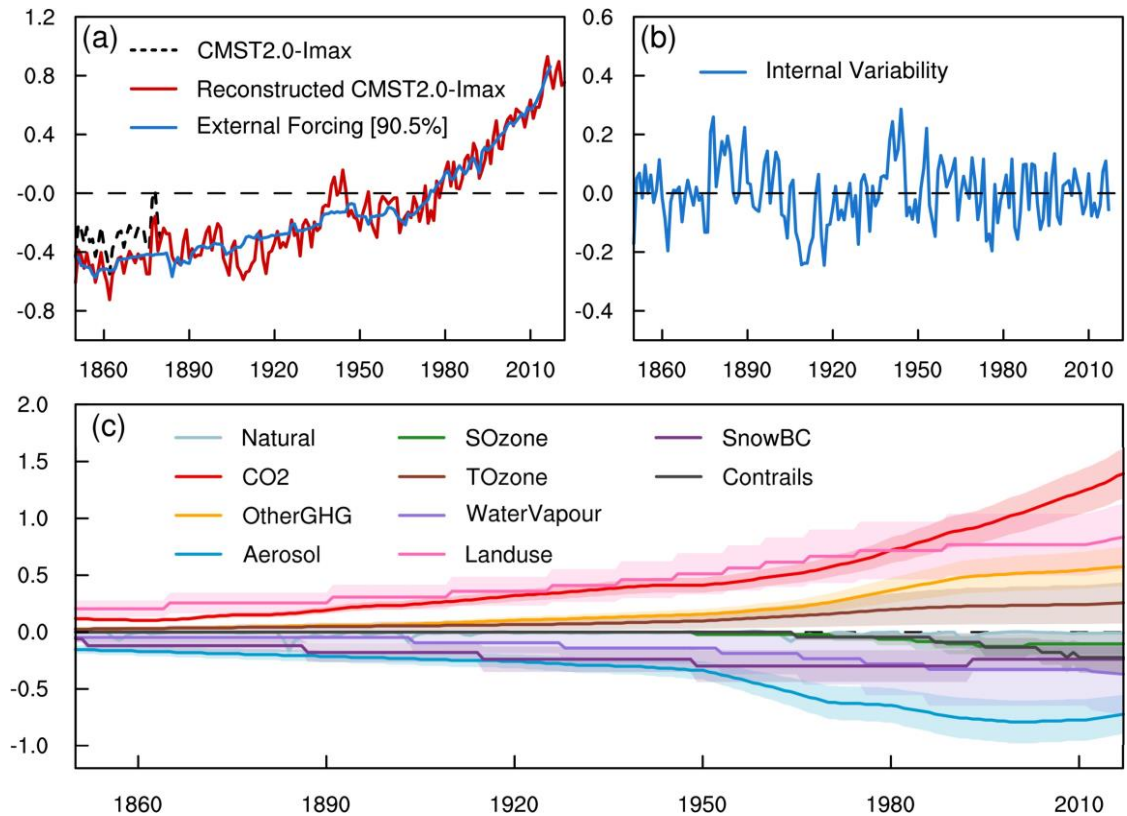

FIG. 3 The reconstruction results of the LEFRM in this study. (a) Original and reconstructed CMST2.0-Imax, and the external forcing (the numbers in square brackets represent the explained variance of the equation); (b) the Internal Variability of the reconstructed CMST2.0-Imax series; (c) the temporal contribution of different forcings (the uncertainties are shown by the range of ±1σ)

*Comparison of the reconstructed GMST series.* —Based on the LEFRM, the global observed ST series mentioned above were reconstructed during 1850-1880 (FIG. 4). Apparently, there is a systematic decrease in the anomalies of the reconstructed series before 1880, indicating that the 1850-1880 anomalies were overestimated by CMST2.0-Imax and CMST2.0-Imin and are corrected and adjusted for after reconstruction (FIG. 4b-c). The warming trends in Table 1 also show good agreement: after reconstruction, the linear trend of CMST2.0-Imax increases from 0.058 ± 0.006 °C/10a to 0.066 ± 0.006 °C/10a; the linear trend of CMST2.0-Imin increased from 0.054 ± 0.006 °C/10a to 0.064 ± 0.006 °C/10a. If the three other datasets (HadCRUT5, NOAAGlobalTemp-Interim, and BE) with data from 1850-1880 are reconstructed in the same way (FIG. 4d-f), their linear trend should also be improved: the reconstructed HadCRUT5 would increase the linear trend from 0.062 ± 0.006°C/10a to 0.069 ± 0.006°C/10a, the reconstructed NOAAGlobalTemp-Interim would increase the linear trend from 0.056 ± 0.006°C/10a to 0.065 ± 0.006°C/10a, and the reconstructed BE from 0.064 ± 0.006°C/10a to 0.069 ± 0.006°C/10a. Adopting the same method to reconstruct the average series for multiple datasets, the external forcing for 1850-2022 reflects a warming trend of 0.063 ± 0.006°C/10a. According to IPCC AR6, compared with the global ST before the period of global industrialization (the average of 1850-1900) in the recent 10 years (2013-2022), the ST rise is about 1.20 ℃ (1.15-1.25 ℃), slightly higher than the 1.09 ℃ given by AR6 (1.07℃, 1.11 ℃, respectively for 2011-2020 and 2013-2022 warming relative to the average of 1850-1900 in this study). That is, the magnitude and rate of current global warming estimation are still slightly underestimated. This is qualitatively consistent with the results of a recent study by Schneider et al.: Based on climate reconstructions from tree-rings, they also noted an overestimation of 19th-century baseline (1850-1900) temperatures [15]. Further, the above results show that based on the reconstruction of global ST during the instrument measurement period in this study, the estimation range of global warming trends since the middle of the 19th century among existing datasets has been further reduced, which shows the uncertainties of global warming estimation have also been further reduced (FIG. 4a). Last but not least, it also suggests the high reliability of the reconstruction method from the high explanation of variance by the regression equation.

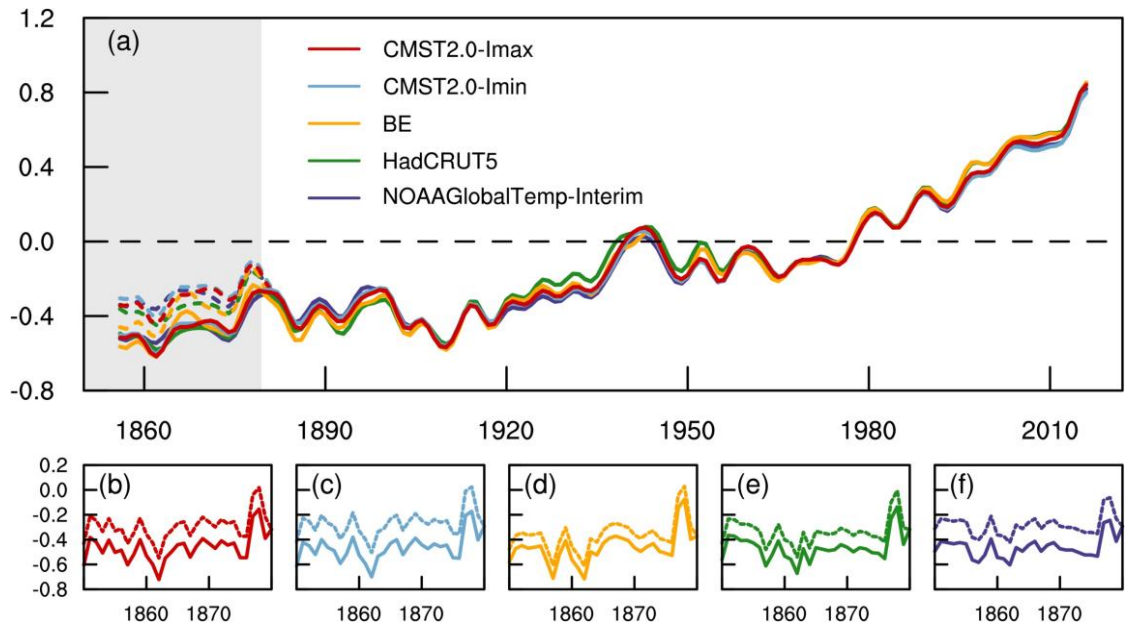

FIG. 4 Reconstructions based on two versions of CMST2.0 (Imax (red) and Imin (blue)), BE (Orange), HadCRUT5 (Green), and NOAAGlobalTemp-Interim (Purple). (a) original (dashed line) and reconstructed (solid line) smoothed GMST series from the above 5 datasets (by a 13-point filter) for 1850-2022; (b-f) original (dashed line) and reconstructed (solid line) annual GMST series from above 5 datasets for 1850-1880.

Table 1. Warming trends (°C/10a) during the period of 1850-2022/1880-2022 from several observational ST datasets used in IPCC AR6 and their reconstructions.

|  | CMST2.0-Imax | CMST2.0-Imin | BE | HadCRUT5 | NOAAGlobT-Int | GISTEMP4 | Average of all datasets |
|---|---|---|---|---|---|---|---|
| **1880-2022** | 0.079±0.008 | 0.076±0.006 | 0.084±0.008 | 0.083±0.008 | 0.078±0.008 | 0.079±0.008 | 0.075±0.008 |
| **1850-2022** | 0.058±0.006 | 0.054±0.006 | 0.064±0.006 | 0.062±0.006 | 0.056±0.010 | —— | —— |
| **Recons 1850-2022** | 0.066±0.006 | 0.064±0.006 | 0.069±0.006 | 0.069±0.006 | 0.065±0.006 | —— | 0.063±0.006 (Forcing only) |

*Outlook.* —We can calculate the linear response of different external forcings as a "deterministic trend" by modeling the GMST with the new statistical physical model (LEFRM). Using this model to simulate the global ST series, we quantitatively extrapolate/predict the external forcing response in the periods of lack of observations, together with IV (interannual, interdecadal, or interannual periodic term), and we can obtain the ST anomalies series for periods of data scarcity. The results show that the reconstructions from different datasets in the 19th century are very consistent and that the temperature anomaly was overestimated for about 0.11°C (0.06~0.16 °C) at this time, which seems to be slightly lower than those of Schneider et al. [15] mentioned above. However, whatever the result, it implies that it is closer to the 1.5°C temperature target set in the Paris Agreement, and the space for future warming will be further compressed.

Actuarial science applies the mathematics of probability and statistics to define, analyze, and solve the financial implications of uncertain future events. When we extend its concept to the field of climate change, it can be regarded as an "actuarial calculation" approach to understanding and predicting global warming rather than the traditional pure statistical reconstruction "evaluation". Based on the above discussion, the authors believe that the era of the "actuarial calculation" of global warming has come, and the accuracy of climate change reconstruction in the global historical period is expected to be further strengthened. It is worth noting that such actuarial calculations are not a negation of the "evaluation" role of previous observations, but rather a mathematical model constraint based on a more "accurate" decomposition of the role of external forcing.

Furthermore, if there are more comprehensive and complete spatial coverage observation and radiative forcing data, our method can also further reconstruct the early ST series at both regional and local scales to obtain more accurate regional and global ST datasets with complete temporal and spatial coverage. Here we do not tend to separate high-frequency IV components in advance, mainly because these components (such as ENSO and volcanoes) may be less stable as they go back to the earlier stages.

We thank the Natural Science Foundation of China (41975105) and the National Key R&D Programs of China (2017YFC1502301, 2018YFC1507705) for financial support.